\documentclass[twocolumn,english,aps,prb,superscriptaddress]{revtex4-1}

\usepackage{amsmath}
\usepackage{graphicx}
\usepackage{amssymb}

\makeatletter

\usepackage{color}

\makeatother

\usepackage{babel}

\begin{document}

\title{
Phase separation in the Edwards model
}

\author{S. Ejima}
\affiliation{Institut f\"ur Physik, Ernst-Moritz-Arndt-Universit\"at Greifswald, 17487 Greifswald, Germany}

\author{S. Sykora}
\affiliation{Institute for Theoretical Solid State Physics, IFW Dresden, 
01069 Dresden, Germany}

\author{K. W. Becker}
\affiliation{Institut f\"ur Theoretische Physik, Technische Universit\"at Dresden, 01062 Dresden, Germany}

\author{H. Fehske}
\affiliation{Institut f\"ur Physik, Ernst-Moritz-Arndt-Universit\"at Greifswald, 17487 Greifswald, Germany}


\begin{abstract}
The nature of charge transport within a correlated background medium
can be described by spinless fermions coupled to bosons 
in the model introduced by Edwards. 
 Combining numerical density matrix renormalization group 
and analytical projector-based renormalization methods  
we explore the ground-state phase diagram of the  
Edwards model in one dimension. 
Below a critical boson frequency any long-range order disappears
and the system becomes metallic. If the charge carriers are coupled to 
slow quantum bosons the Tomonaga-Luttinger liquid is attractive 
and finally makes room for a phase separated state, just as in the $t$-$J$ 
model. The phase boundary separating the repulsive from the attractive 
Tomonaga-Luttinger liquid is determined from long-wavelength 
charge correlations, whereas fermion segregation is indicated by a 
vanishing inverse compressibility. 
On approaching phase separation the photoemission spectra 
develop strong anomalies.
\end{abstract}

\date{\today}


\maketitle
\section{Problem}
The Edwards fermion-boson model\cite{Ed06,EEAF10} constitutes a paradigmatic 
model for the theoretical description of quantum transport in solids. 
Charge transport normally takes place in the presence of some background 
medium.\cite{Be09} Examples for such a ''background'' could be a 
spin-, charge- or 
orbital-ordered lattice,\cite{WL02,WOH09} but also a sequence of chemical side 
groups along the transport path, a deformable medium, or even a heat bath might be possible.  In all these cases, the transfer of the charge carriers will  
be strongly influenced by fluctuations, 
which may exist in the background medium. 
The other way around, the properties of the background are quite often 
determined by the motion of the particle itself. 

Correlations inherent in such a complex interactive system 
are mimicked  by a boson-affected
hopping of spinless fermionic particles in the Edwards model. It reads:  
\begin{equation}
 H = -t_b\sum_{\langle i, j \rangle} f_j^{\dagger}f_{i}^{}
  (b_i^{\dagger}+b_j^{})- \lambda\sum_i(b_i^{\dagger}+b_i^{})
              + \omega_0\sum_i b_i^{\dagger}b_i^{}\,.
\label{model}
\end{equation}
Every time a fermion $f_i^{(\dagger)}$ hops, it creates or absorbs  
a boson $b_i^{(\dagger)}$ of energy $\omega_0$ at the sites it 
leaves or enters. Such an excitation
or de-excitation corresponds to a local ``distortion'' of the background. 
Because of quantum fluctuations
the distortions are able to relax ($\propto \lambda$).   
The physically most interesting regimes in this setting are those of vanishing 
fermion density and of a  half-filled band, representing 
doped Mott insulators, polaronic organics  
and charge-density-wave (CDW) systems\cite{AEF07,WFAE08,EHF09,EF09b} 
with possible relevance to high-$T_c$ superconductors,\cite{BM86,Da94} 
colossal magnetoresistive manganites,\cite{JS50,Da03,WF04b} 
carbon nanotubes,\cite{SDD98,MB12} 
graphene\cite{NGMJZDGF04,SP08} and CDW transition 
metal complexes,\cite{Cl84,BB87} respectively. However,   
the Edwards model also reveals fascinating properties over the 
whole density range.

On these grounds, the main goal of the present work is to pinpoint 
the ground-state phase diagram of the 
one-dimensional (1D) Edwards model as a function of the
band filling $n$.  
Thereby we demonstrate that this model indeed captures a number  
of very interesting phenomena, including e.g.~electronic phase separation (PS). 
To obtain reliable information about the ground-state and spectral
properties of the model in the thermodynamic limit, we employ numerical  
pseudosite density-matrix renormalization group (DMRG)
and dynamical DMRG (DDMRG) techniques (supplemented by a careful 
finite-size scaling procedure, for details 
see Refs.~\onlinecite{Wh92,JW98b,Je02b,JF07,EF09b}), in combination with the 
analytical projective renormalization method (PRM).\cite{BHS02,SBF10} 

So far the 1D Edwards model has been
solved exactly by numerical approaches for two cases.\cite{BTB99,JW98b,JF07}  
The first, in no way trivial case concerns just a single particle on  
an infinite lattice, where---depending on the model parameters---transport 
appears to be quasi-free, diffusive, or boson-assisted.\cite{AEF07,AEF10}  
When strong correlations develop at small $\lambda$ and large $\omega_0$,     
the background becomes ``stiff'' and the spinless particle's motion 
resembles those of a hole in an antiferromagnetic 
spin background,\cite{KLR89,MH91a} as known from 
the $t$-$J$ model. Interestingly, the Edwards
model allows for so-called Trugman paths\cite{Tr88} in an 1D setting
(Trugman paths usually unwind the string of misaligned spins 
a mobile hole leaves behind in two dimensions). 
The second case is half-filling $n=1/2$.  
Here a metal-insulator transition has been proven to exist: 
For small $\lambda$ and large $\omega_0$ the repulsive Tomonaga-Luttinger 
liquid (TLL) gives way to a CDW.\cite{WFAE08,EHF09} 
The CDW constitutes a few-boson state that typifies a Mott-Hubbard 
insulator rather than a Peierls state (normally established by the softening 
of a boson mode).  
\section{Theoretical Results}
\subsection{Ground-state properties}
The situation at finite density $n$ ($n\neq 1/2$) is much 
less understood. 
By analogy with the $t$-$J$ model one might expect that the system 
is metallic for $0<n<1/2$, at least if the background is 
not too stiff. If so, the next question will be, whether the Edwards 
model might support the pairing of electrons in a certain parameter 
regime $(\lambda, \omega_0)$. Apparently a second electron, 
following the path of a first one, can take advantage of the 
background excitations (bosons) left behind by the first
electron. This acts like an effective attractive interaction. If this  
attraction completely dominates the kinetic energy the  
system might even phase-separate into particle-enriched and 
particle-depleted regions.\cite{EKL90} Since the Edwards model 
is not particle-hole symmetric, it is a moot point whether 
a corresponding hole pairing mechanism is at play also for $1/2<n<1$.

To address these problems, we analyze the charge correlations
existing in the ground-state of the 1D Edwards model. Firstly, we   
calculate the charge structure factor    
\begin{equation}
 S_c(q)=\frac{1}{L}\sum_{j,l}e^{iq(j-l)}
    \left\langle \left(f^\dagger_{j}f^{\phantom{\dagger}}_{j} -n\right) 
    \left(f^\dagger_{l}f^{\phantom{\dagger}}_{l}-n\right)\right\rangle\,,
\label{csf}
\end{equation}
for a system with $N$ fermions, $L$ sites, and open boundary conditions (OBC).
So the particle density is $n=N/L$ and the momenta $q=2\pi s /L$
with integers $0< s< L$.
The TLL charge exponent is proportional to the slope 
of $S_c(q)$ in the long-wavelength limit $q\to 0^+$ (cf. Refs.~\onlinecite{Dz95,EGN05}):
\begin{equation}
 K_\rho=\pi\lim_{q\to 0}\frac{S_c(q)}{q}\,.
\label{kro}
\end{equation} 
Then,
$K_\rho>1$ ($K_\rho<1$) characterizes an attractive (repulsive) 
TLL  for our spinless fermion transport model~\eqref{model}, 
and $K_\rho=1/2$ will define a metal-insulator transition 
point at $n=1/2$.\cite{Gi03} That the finite-size scaling 
of $S_c(q)$ and $K_\rho$ works well has been demonstrated for the 
half-filled band case.\cite{EHF09} 
Secondly, in order to comprise potential PS, we  
 determine the finite-size equivalent of the charge 
compressibility,~\cite{OLSA91}
\begin{equation}
 \kappa=\frac{L}{N^2}\left[\frac{E_0(N+2)+E_0(N-2)-2E_0(N)}{4}\right]^{-1}\,,
\label{kappa}
\end{equation}
with $E_0(N)$ being the ground-state energy for $N$ electrons on $L$ sites.  
An infinite compressibility  signals the formation of a PS state.
\begin{figure}[h]
   \centering
    \includegraphics[clip,width=0.48\linewidth]{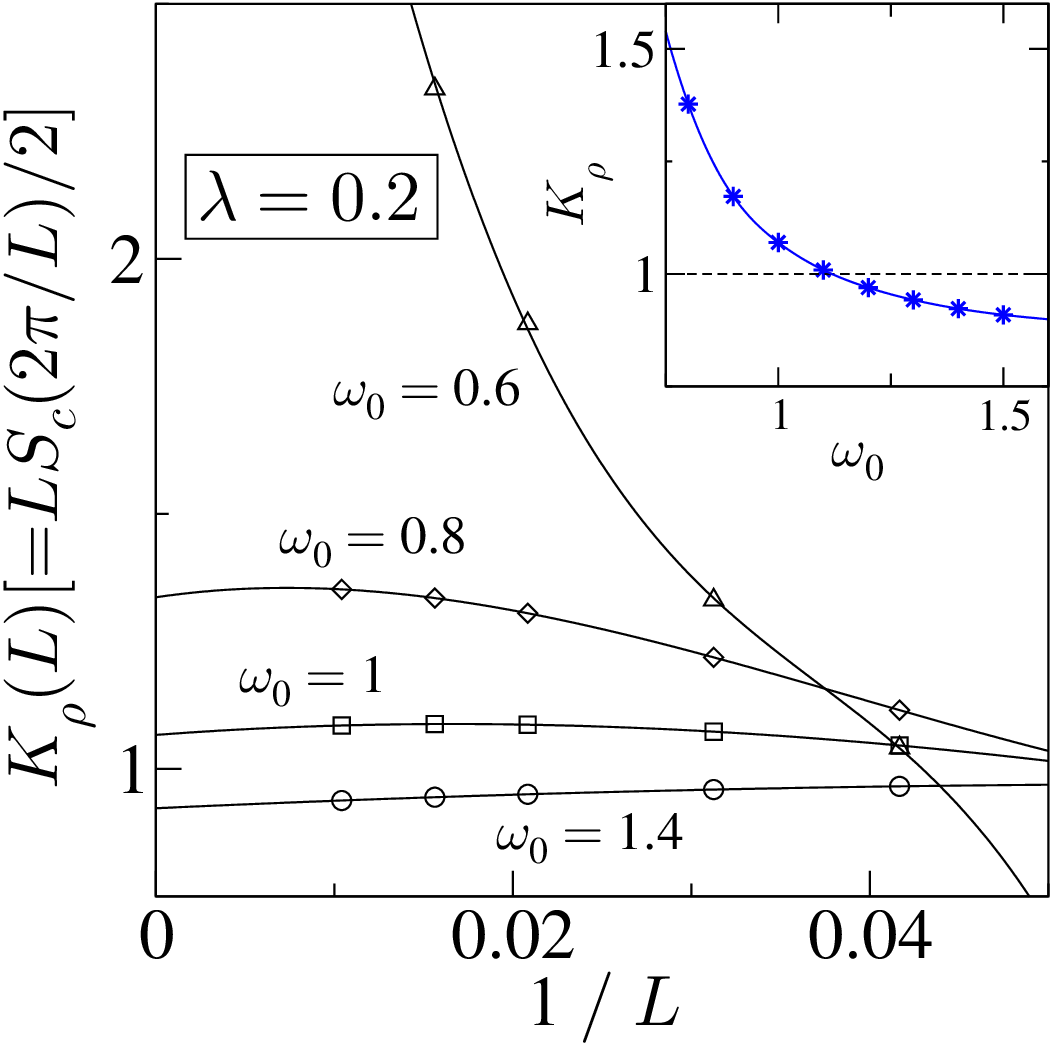}
    \hspace*{0.2cm}\includegraphics[clip,width=0.48\linewidth]{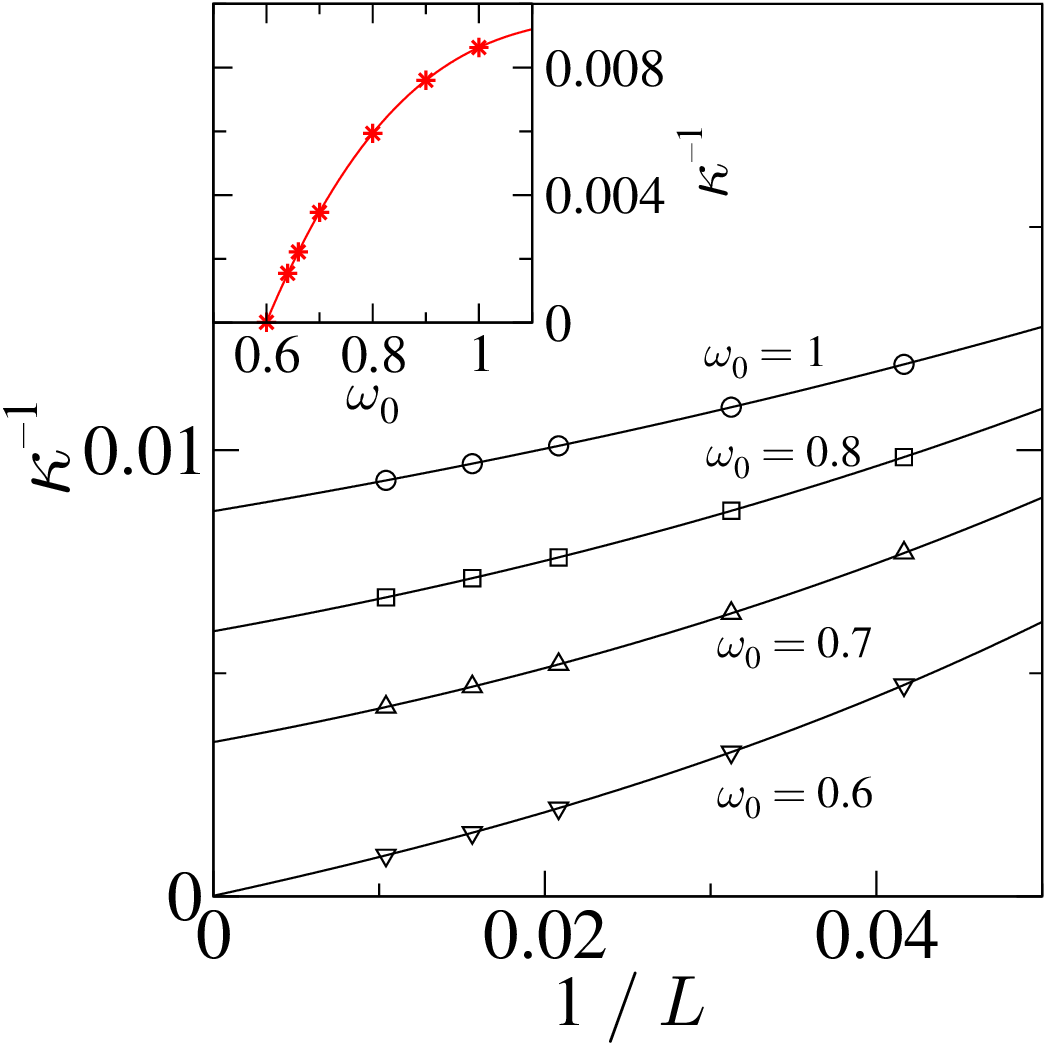}
\caption{Finite-size scaling of the TLL parameter $K_\rho$ (left panel) 
and inverse compressibility $\kappa^{-1}$ (right panel) for the 1D 
Edwards model with $\lambda=0.2$ at $n=1/8$. Insets show the 
variation with $\omega_0$ of the $L\to\infty$ extrapolated values.
Results are obtained by DMRG for a lattice with $L$ 
sites and OBC. In all DMRG runs we take into account up to $n_b = 4$
bosonic pseudosites and determine $n_b$ by the requirement that the local
boson density of the last pseudosite is less than $10^{-6}$
for all $j$. Using selectively $n_b=5$, we controlled that 
convergence is truly achieved. Furthermore we keep up to $m = 1200$
density-matrix eigenstates in the renormalization steps
to ensure that the discarded weight is smaller than $10^{-8}$. 
}
\label{fig:finite-size-scaling}
\end{figure} 

Fig.~\ref{fig:finite-size-scaling} demonstrates that all this works 
in practice. Exemplarily choosing $n=1/8$ and $\lambda=0.2$, we show
how the TLL exponent $K_\rho$ and the inverse compressibility 
$\kappa^{-1}$ scale with increasing system size at various
$\omega_0$. The transition point between a repulsive TLL---obviously 
realized at 
$\omega_0>\omega_{0,c1}$ when excitations of the background 
are energetically rather costly---and an attractive TLL 
at smaller $\omega_0$, can be read off from the inset [depicting the 
extrapolations $K_\rho(L\to\infty)$] to be 
$\omega_{0,c1}(\lambda= 0.2)\simeq 1.118$. 
Reducing $\omega_0$ further, in the attractive TLL phase, 
a dramatic increase in $K_\rho$ is observed at $\omega_{0,c2}\simeq 0.6$.
Our inverse compressibility data indicate that at this point the attraction 
among the particles becomes so strong that the system shows PS, i.e.,
$\kappa^{-1}= 0$ for $\omega_{0}<\omega_{0,c2}$.

\begin{figure}[t]
 \centering
  \includegraphics[clip,width=\linewidth]{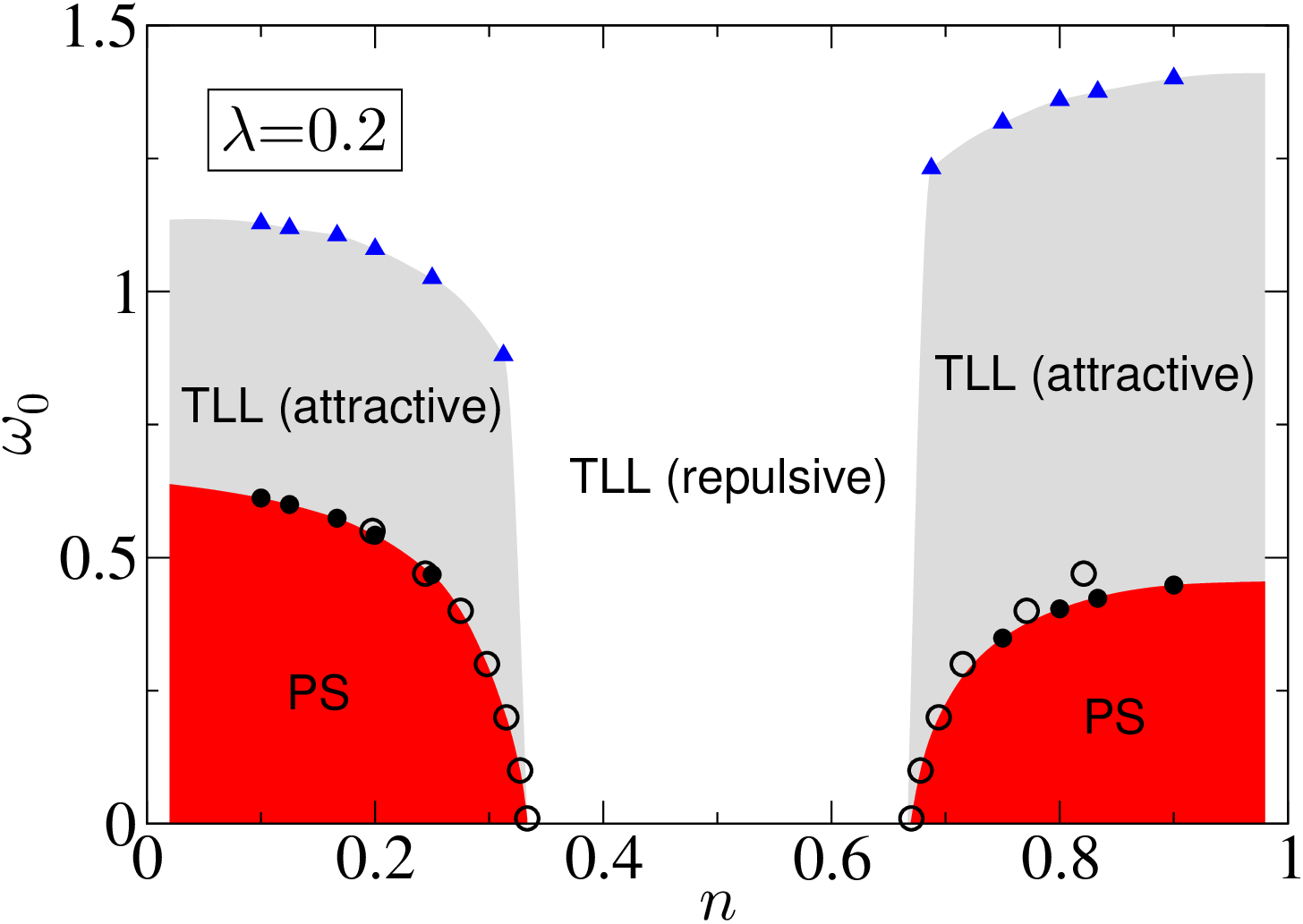}\\[0.1cm]
  \includegraphics[clip,width=\linewidth]{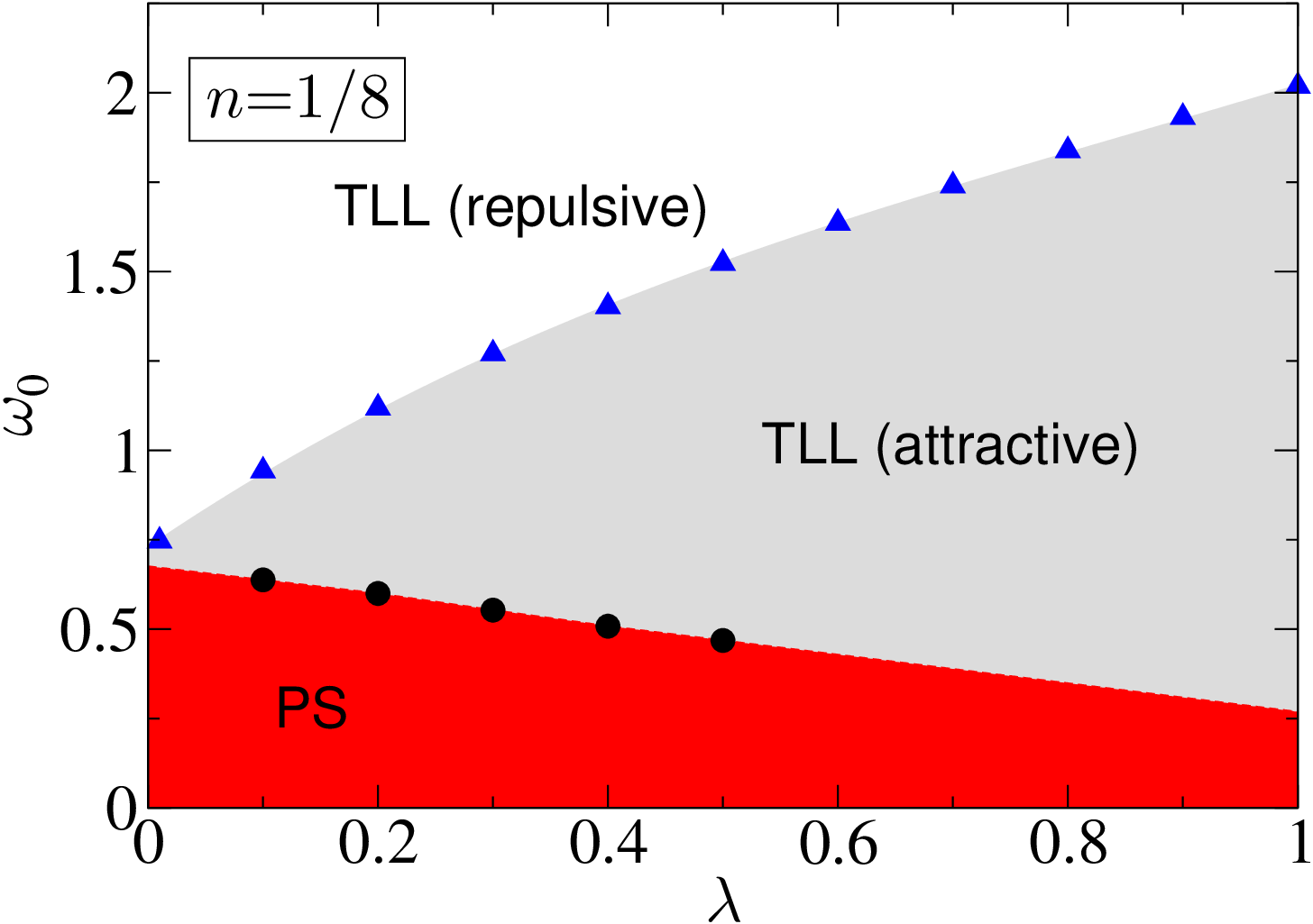}
  \caption{Ground-state phase diagram of the 1D Edwards model, showing
repulsive TLL ($K_\rho<1$), attractive TLL ($K_\rho>1$), and 
PS ($\kappa^{-1}= 0$) regions. The phase boundaries were 
obtained by DMRG (filled symbols) in the course of a 
careful finite-size scaling analysis and for the infinite system directly 
by PRM (open symbols). The upper panel displays
the phase diagram as a function of $n$, varying $\omega_0$ 
at fixed $\lambda=0.2$; the lower panel gives the phase diagram
in the $\lambda$-$\omega_0$ plane for fixed density $n=1/8$.}
\label{fig:pd}
\end{figure}

Proceeding in the same manner for different values of $n$ and $\lambda$
respectively $\omega_0$, we can map out the phase diagram of the 1D 
Edwards model. The outcome is displayed in  Fig.~\ref{fig:pd}.
Let us first consider the case of a not too small boson relaxation 
$(\propto\lambda)$, which  ensures that the system 
is metallic for large $\omega_0$ in the whole density regime. 
Then, as the upper panel
of Fig.~\ref{fig:pd} shows, depending on $n$ we find up to three different regimes: For small and large particle densities a repulsive TLL, an attractive 
TLL and a PS state appear in sequence as the energy of the bosons
is lowered. In contrast, around half-filling only the repulsive TLL exists. 
At this point we would like to remind the readers that at half-filling, 
for smaller values of  $\lambda$, 
($ \lambda < \lambda_c$), a CDW is formed 
for $\omega_0 >\omega_{0,c}$.\cite{EHF09} 
Our unbiased DMRG calculations give no evidence for any other phases.   
Note that the behavior in the 
low-density regime is consistent with what is found for the 1D $t$-$J$
model\cite{OLSA91,MMM11}, where the holes correspond to the spinless fermions 
in the Edwards model.  
As mentioned before, the phase diagram is not symmetric with respect to
$n=1/2$. This is because the hopping of a hole 
(missing electron) to a neighboring 
site creates in the Edwards model a boson at the arrival site and not 
on the departure site as in the motion of a single particle. Since
this boson can be destroyed immediately when the hole makes a further hop,
there is no string effect and a few holes propagate more easily than 
a few electrons. Hence the background at fixed $\omega_0$
and $\lambda$ appears to be less stiff for $n\lesssim 1$ than for
$n\gtrsim 0$ and, as a result, the boundary between the repulsive and 
attractive TLL is shifted to larger values of $\omega_{0,c1}$. Contrariwise,
since carriers will be less mobile in a state with charge separation, 
the boundary between the attractive TLL and the PS state 
$\omega_{0,c2}$ is shifted to smaller values if one compares the 
corresponding results at high and low carrier density.     
   
We next consider a fixed particle density $n=1/8$ and track the
phase boundaries in the $\lambda$-$\omega_0$ plane; see the lower
panel of Fig.~\ref{fig:pd}. The repulsive TLL established for large 
$\omega_0$ is strongly renormalized if the 
ability of the background to relax is low. 
For example, we find $K_\rho=0.527$ for $\omega_0=2$ at $\lambda=0.01$.
Lowering $\omega_0$ for such a stiff background, the transition to the 
PS state happens almost instantaneously with a narrow
intervening attractive TLL state; at $\lambda=0.01$ we have 
$\omega_{0,c1}\simeq 0.747$ and $\omega_{0,c2}\simeq 0.672$. If we fix,
on the other hand, $\omega_0=2$ and increase $\lambda$, we observe
a strong enhancement of $K_\rho$ in the interval  $0<\lambda<0.3$
($K_\rho=0.767$, 0.879, 0.932 for $\lambda=0.1$, 0.2, 0.3, respectively),
followed by a very gradual increase until the transition to the attractive
TLL takes place at about $\lambda_{c1}\simeq 1$. Obviously, the 
region where the attractive TLL (PS state) 
constitutes the ground state expands (shrinks) as $\lambda$
gets larger, which can be traced back to the subtle competition between
kinetic energy and charge segregation effects.

\begin{figure}
 \begin{center}
 \includegraphics[clip,width=\columnwidth]{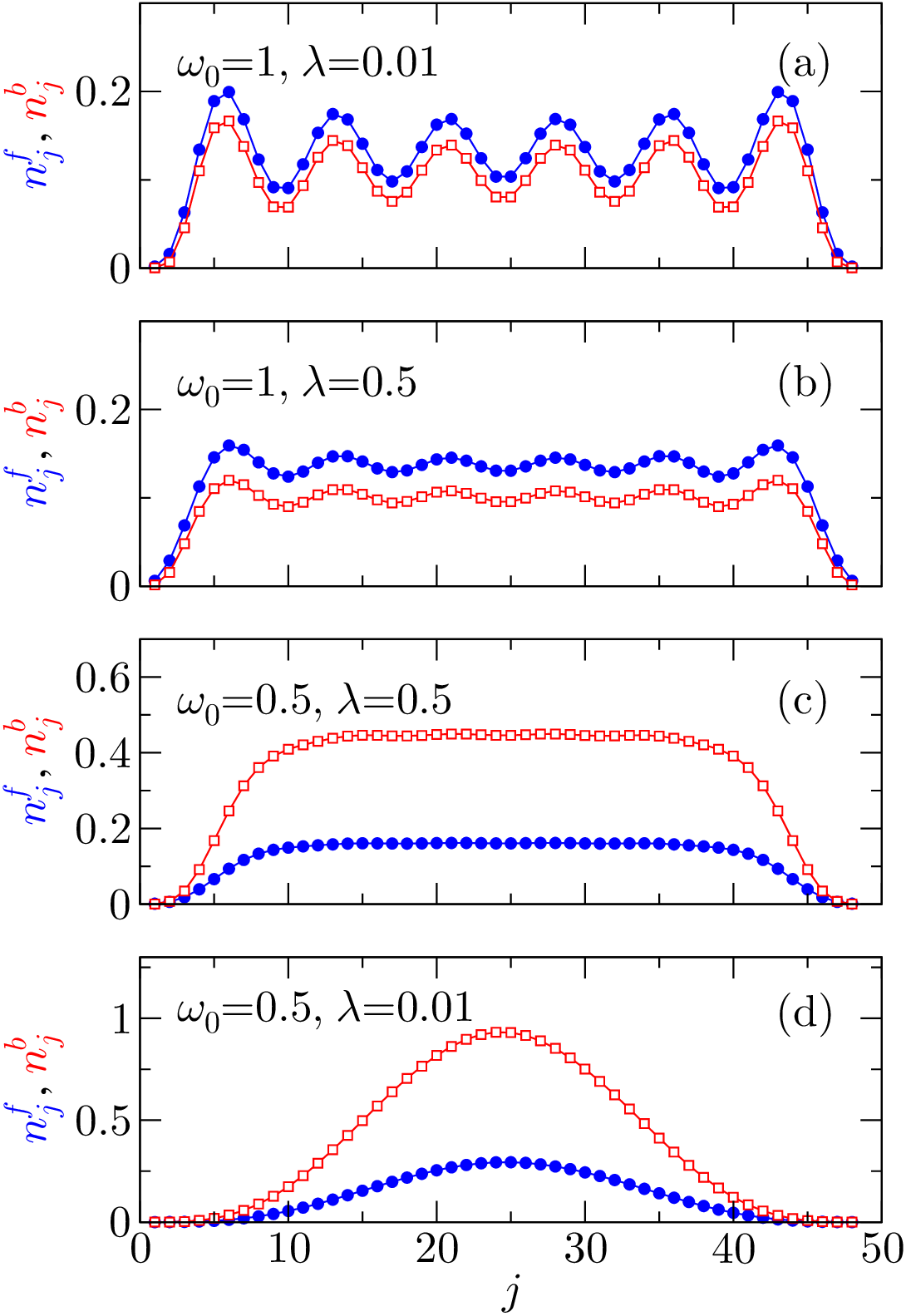}
 \end{center}
 \caption{Local particle (filled circles) and boson (open squares) densities for the 
1D Edwards model with $n=1/8$, calculated for a
48-site system with OBC at various $\lambda$ and $\omega_0$.}
 \label{ldens}
\end{figure}
To gain deeper insight into the different mechanisms at play, 
we analyze the variation of the local fermion 
[$n_j^f =\langle f_j^\dagger f_j^{}\rangle$] and boson 
[$n_j^b =\langle b_j^\dagger b_j^{}\rangle $] densities
in real space. In Fig.~\ref{ldens} we show 
both density profiles for $n=1/8$ and characteristic
values of $\lambda$ and $\omega_0$, implementing  
repulsive TLL, attractive TLL, and PS states. 
In the metallic regime, the OBC lead to Friedel oscillations 
in the particle density. These oscillations are especially 
pronounced for the repulsive TLL [panel (a)], and they will be even 
stronger for $\omega_0=2$ (not shown), where the number
and fluctuations of the bosons is reduced.
Note that the Friedel oscillations caused by the OBC 
will be algebraically reduced, if we move for very large 
systems towards the central part of the chain. If we 
enter the attractive TLL regime by increasing $\lambda$ at 
fixed $\omega_0$ the Friedel oscillation will be smeared out
[panel (b)]. Thereby, the number of oscillations stays constant,  
which means that a pairing of electrons does not occur.
This is an important difference from the spinful 1D $t$-$J$ model,  where
a recent DMRG study reveals that the number of Friedel oscillations is halved
(by increasing $J$), corresponding to half the number of particles, 
which indicates pairing.\cite{MMM11} Next lowering $\omega_0$ to a point 
where PS sets in, the particle density oscillations vanish, 
see panel (c). There,  just above the PS boundary, no evidence is found
for the clustering of multiple particles in several groups. This agrees with the
findings for the 1D $t$-$J$ model.\cite{MMM11} Once we are deep inside
the PS region, a density distribution results, where
a single island of particles at the central part of the system appears, 
leaving a sizable number of almost empty sites at both ends of the chain 
[see panel (d)]. Looking at the bosonic degrees of freedom, we
see that the strong attraction among these particles is mediated
by a boson cloud covering the electron-rich region. As a result
the kinetic fluctuations will be strongly quenched.  

It is all but impossible to comprise by DMRG the large number of bosons and the 
strong bosonic fluctuations, which  appear at still smaller $\omega_0$  in the 
Edwards model, simply because the dimension of the Hilbert space 
increases dramatically. To access the low-$\omega_0$ regime 
and reconfirm the DMRG phase boundaries, obtained for the 1D Edwards model 
at larger values of $\omega_0$, we employ the analytical PRM 
approach.\cite{BHS02} The basic idea of the PRM is to construct---performing 
a sequence of discrete unitary transformations and eliminating
all transitions with energy larger than a given cutoff energy---an 
effective non-interacting Hamiltonian $\tilde{H}$ with renormalized 
parameters (in the limit of vanishing cutoff energy). For the metallic 
state of the Edwards model, in momentum space, it takes the form 
\begin{equation}
\tilde{H} = \sum_k \tilde{\varepsilon}_k f_k^\dag f_k^{} + 
\sum_q \tilde{\omega}_q b_q^\dag b_q^{} + \tilde{E}\,,
\end{equation}
where the renormalization equations 
for $\tilde{\varepsilon}_k$ and  $\tilde{\omega}_q$ 
have been derived in Ref.~\onlinecite{SBF10} (for the half-filled band case). 
In order to fix the boundary between metallic and PS 
states,  we (i) solve the renormalization equations at a given $E_F$ in the 
TLL phase, (ii) determine the corresponding fermion density $n$,
and (iii) slightly vary $E_F$ 
(to get closer to the PS instability),
determine $n$, and repeat the whole procedure self-consistently
until the functional dependence of $n$ on $E_F$ is established. A vanishing
inverse compressibility can then be read off from a sudden jump of $n$ under
a tiny variation of $E_F$ or, the other way around, if a plateau    
in the $E_F$ versus $n$ plot appears. The transition points, determined 
in the manner described, were inserted in Fig.~\ref{fig:pd} to complete
the ground-state phase diagram in the $n$-$\omega_0$ plane. Whenever
points can be compared we find a remarkable agreement between our
DMRG and PRM data. The deviation of the PRM data at larger particle 
densities $n\gtrsim 0.75$ results from uncertainties in fixing the jump of 
$n$ under $E_F$ variation. Here we should trust in the DMRG phase boundary.
\subsection{Spectral properties}
Let us finally discuss the single-particle spectrum of the 1D Edwards
model. The single particle excitations associated with the 
injection ($+$) or emission ($-$) of an electron with wave vector $k$, 
\begin{equation}
 A^{\pm}(k,\omega)
=\sum_m|\langle\psi_m^{\pm}|f^{\pm}_k|\psi_0\rangle|^2\,
\delta (\omega\mp\omega^{\pm})\,,
\label{spsfpm}
\end{equation}
can be computed by DDMRG,\cite{Je02b,JF07} 
where $f^+_k=f^\dagger_k$, $f^-_k=f^{}_k$, $|\psi_0\rangle$
is the ground state of a $L$-site system in the $N$-particle sector,
and $| \psi_m^{\pm}\rangle$ denote the $m$-th
excited states in the $(N\pm 1)$-particle sectors
with excitation energies $\omega_m^\pm=E_m^\pm-E_0$.

Within PRM  we find for the photoemission part 
\begin{eqnarray}
A^-(k,\omega) &=& \;\tilde{\alpha}_k^2 \, \tilde{n}^f_k
\, \delta(\omega - \tilde{\varepsilon}_k) \nonumber\\ &&+ \sum_q \tilde\beta_{k,q}^2 \,
( 1 + \tilde{n}^b_q) 
\tilde{n}^f_{k+q} 
\delta(\omega + \tilde{\omega}_q - \tilde{\varepsilon}_{k+q}) 
\nonumber\\ &&+ \sum_q \tilde\gamma_{k,q}^2 \,
\tilde{n}^b_q \tilde{n}^f_{k-q}
\delta(\omega - \tilde{\omega}_q - \tilde{\varepsilon}_{k-q})\,, 
\label{spsfprm}
\end{eqnarray}
where $\tilde{n}^f_q$ ($\tilde{n}^b_q$) are the fermion (boson) 
occupation numbers in momentum space calculated with the 
renormalized Hamiltonian $\tilde{H}$.  
The coefficients $\tilde{\alpha}_k^2$, $\tilde\beta_{k,q}^2$, 
and  $\tilde\gamma_{k,q}^2$
follow from renormalization equations.\cite{SBF10} 
Taking the corresponding expression for $A^+(k,\omega)$, the sum rule 
$
\int_{\infty}^\infty d\omega [A^+(k,\omega)+A^-(k,\omega)]=1$ 
is fulfilled.
\begin{figure}[b]
\centering
\includegraphics[clip,width=\columnwidth]{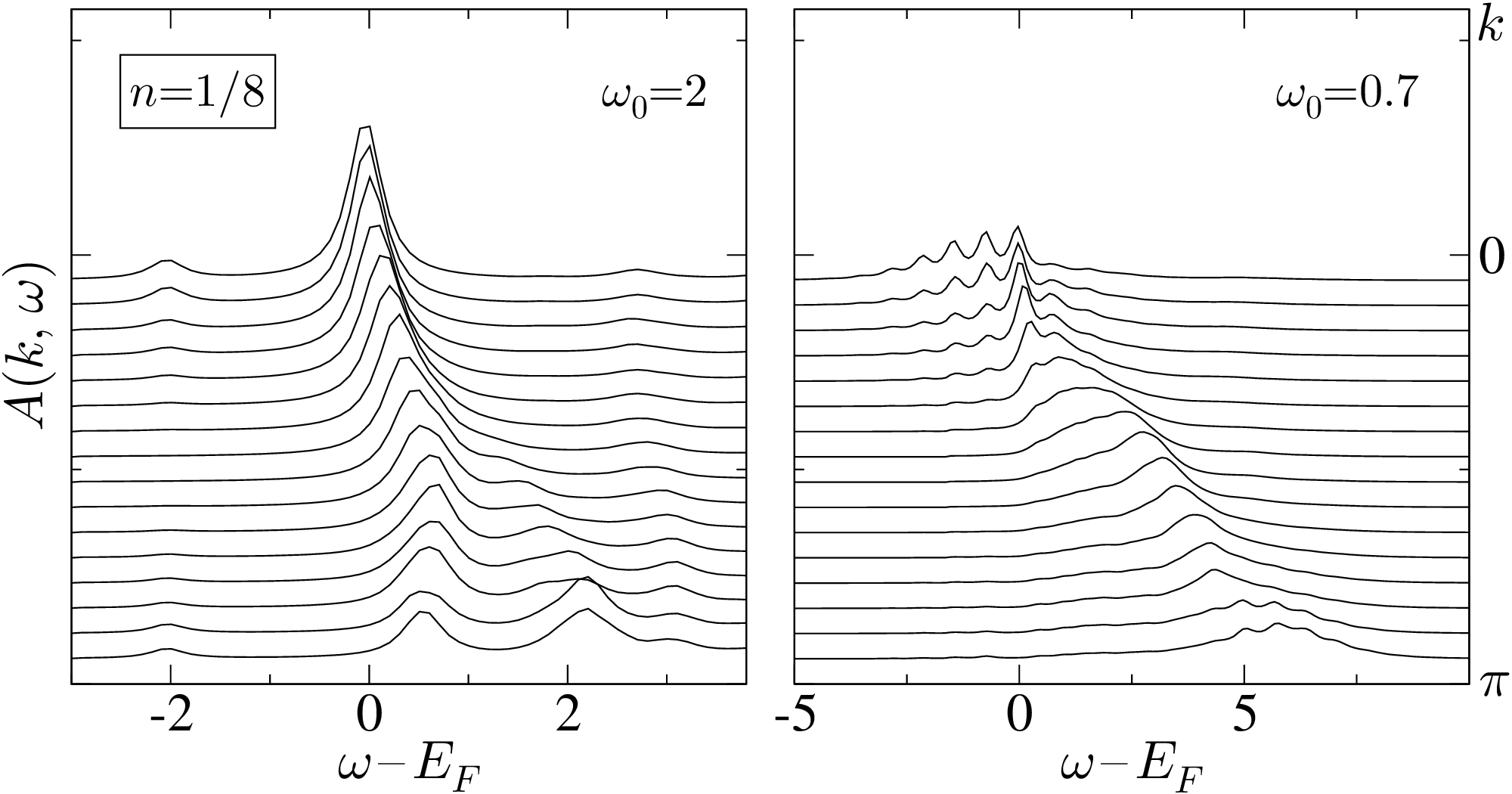}
 \caption{Line-shape of the single-particle spectral function $A(k,\omega)$
for the 1D Edwards model with $\lambda=0.2$ at $n=1/8$.
Results are obtained by DDMRG  for an 16-site chain with quasimomenta
$k=\pi s/(L+1)$, using a broadening $\eta=0.2$.}
 \label{Akw-DDMRG}
\end{figure}

\begin{figure}
 \centering
 \includegraphics[clip,width=\columnwidth]{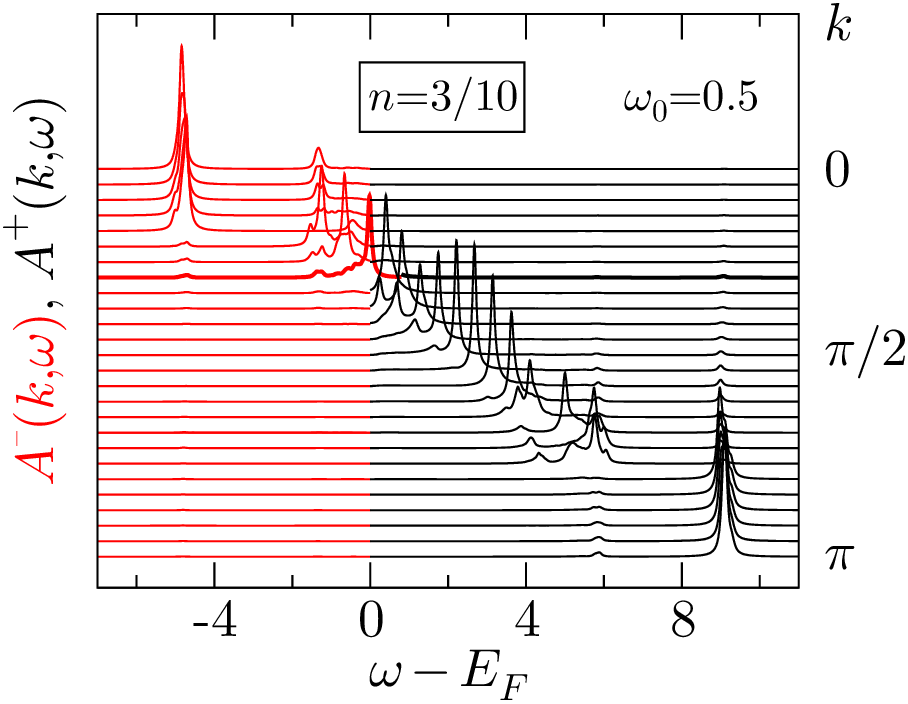}
 \caption{PRM [inverse] photoemisson spectrum [$A^{+}(k,\omega)$] 
$A^{-}(k,\omega)$ for the Edwards model with $\lambda =0.2$, $n=3/10$.
The bold line marks the signal at the Fermi momentum $k_F=0.3\pi$.}
 \label{Akw-PRM}
\end{figure}

Figure~\ref{Akw-DDMRG} gives the combined photoemission spectrum, 
$A(k,\omega)=A^+(k,\omega)+A^-(k,\omega)$, as obtained
by DDMRG  for quasimomenta $k$. In the repulsive TLL regime
(left panel, $\omega_0=2$) we find a rather coherent 
signal with comparable spectral
weight for all $k$ values $k\lesssim \pi/2$. For larger $k$, excitations with at least
one additional $\omega_0$-boson involved become important 
(recall that both initial $N$-particle- and 
target $(N\pm 1)$-particle states are multiboson states with a 
momentum being the total momentum of electrons and bosons).   
The spectrum in the attractive TLL phase (right panel, $\omega_0=0.7$) shows
a sharp absorption signal in the vicinity of $k_F$ only.
Here the (inverse) photoemission spectrum for $k<k_F$  ($k>k_F$)
exhibits a few absorption maxima at multiples of the boson energy.
Obviously, due to the smaller $\omega_0$, here the dynamics of the system
becomes dominated by bosonic fluctuations.

The (inverse) photoemission close to the transition to the PS state
is depicted Fig.~\ref{Akw-PRM}, as calculated by PRM for $\omega_0=0.5$.
Here the almost dispersionless signal for $k\leq 0.212 \pi$ 
($k\geq 0.788 \pi$) is a precursor of the PS instability.   As we noted  
above, $E_F$ does not show any variation with $n$ if $n$ is smaller (larger)
than the lower (upper) critical density $n_{c2}$ for PS. 
Since in the PRM derivation of $A^{\pm}(k,\omega)$ terms with two 
(and more) bosonic creation or annihilation operators were 
neglected,\cite{SBF10} the photoemission spectrum of Fig.~\ref{Akw-PRM}
does not feature the multiboson related signatures found within   
a DDMRG treatment.  
\section{Conclusions}
To summarize, the combination of analytical (PRM) and numerical (DMRG)
approaches permits the precise determination of the ground-state
phase diagram of the 1D Edwards model in the whole parameter regime. 
In the low- and high-density regions, the 
attraction between the particles mediated by the bosonic degrees of freedom 
representing the background medium might become so strong that electronic
PS sets in. In the remaining region the system realizes a TLL or, at $n=1/2$, 
possibly even a truly long-range ordered CDW state. Depending on the
properties of the background medium, the TLL might be attractive
or repulsive. The richness of the phase diagram is remarkable. 
The model captures important features of Holstein, $t$-$J$, Hubbard,
and Falicov-Kimball type models.\cite{AEF07,WFAE08,EHF09,EF09b,BF10}   
Since the Edwards model is one of the simplest models for 
studying transport in low-dimensional systems,
an inspection of its predictions with ultracold fermion/boson 
quantum gases,\cite{BDZ08} as, e.g., carried out for the 1D $t$-$J$ 
model,\cite{EL10} would be of great interest.
\acknowledgements
This work was supported by DFG through SFB 652.

\bibliography{ref} 

\begin{thebibliography}{40}
\expandafter\ifx\csname natexlab\endcsname\relax\def\natexlab#1{#1}\fi
\expandafter\ifx\csname bibnamefont\endcsname\relax
  \def\bibnamefont#1{#1}\fi
\expandafter\ifx\csname bibfnamefont\endcsname\relax
  \def\bibfnamefont#1{#1}\fi
\expandafter\ifx\csname citenamefont\endcsname\relax
  \def\citenamefont#1{#1}\fi
\expandafter\ifx\csname url\endcsname\relax
  \def\url#1{\texttt{#1}}\fi
\expandafter\ifx\csname urlprefix\endcsname\relax\def\urlprefix{URL }\fi
\providecommand{\bibinfo}[2]{#2}
\providecommand{\eprint}[2][]{\url{#2}}

\bibitem[{\citenamefont{Edwards}(2006)}]{Ed06}
\bibinfo{author}{\bibfnamefont{D.~M.} \bibnamefont{Edwards}},
  \bibinfo{journal}{Physica B} \textbf{\bibinfo{volume}{378-380}},
  \bibinfo{pages}{133} (\bibinfo{year}{2006}).

\bibitem[{\citenamefont{Edwards et~al.}(2010)\citenamefont{Edwards, Ejima,
  Alvermann, and Fehske}}]{EEAF10}
\bibinfo{author}{\bibfnamefont{D.~M.} \bibnamefont{Edwards}},
  \bibinfo{author}{\bibfnamefont{S.}~\bibnamefont{Ejima}},
  \bibinfo{author}{\bibfnamefont{A.}~\bibnamefont{Alvermann}},
  \bibnamefont{and} \bibinfo{author}{\bibfnamefont{H.}~\bibnamefont{Fehske}},
  \bibinfo{journal}{J. Phys. Condens. Matter} \textbf{\bibinfo{volume}{22}},
  \bibinfo{pages}{435601} (\bibinfo{year}{2010}).

\bibitem[{\citenamefont{Berciu}(2009)}]{Be09}
\bibinfo{author}{\bibfnamefont{M.}~\bibnamefont{Berciu}},
  \bibinfo{journal}{Physics} \textbf{\bibinfo{volume}{2}}, \bibinfo{pages}{55}
  (\bibinfo{year}{2009}).

\bibitem[{\citenamefont{Wegener and Littlewood}(2002)}]{WL02}
\bibinfo{author}{\bibfnamefont{L.~G.~L.} \bibnamefont{Wegener}}
  \bibnamefont{and} \bibinfo{author}{\bibfnamefont{P.~B.}
  \bibnamefont{Littlewood}}, \bibinfo{journal}{Phys. Rev. B}
  \textbf{\bibinfo{volume}{66}}, \bibinfo{pages}{224402}
  (\bibinfo{year}{2002}).

\bibitem[{\citenamefont{Wohlfeld et~al.}(2009)\citenamefont{Wohlfeld, Ole\'{s},
  and Horsch}}]{WOH09}
\bibinfo{author}{\bibfnamefont{K.}~\bibnamefont{Wohlfeld}},
  \bibinfo{author}{\bibfnamefont{A.~M.} \bibnamefont{Ole\'{s}}},
  \bibnamefont{and} \bibinfo{author}{\bibfnamefont{P.}~\bibnamefont{Horsch}},
  \bibinfo{journal}{Phys. Rev. B} \textbf{\bibinfo{volume}{79}},
  \bibinfo{pages}{224433} (\bibinfo{year}{2009}).

\bibitem[{\citenamefont{Alvermann et~al.}(2007)\citenamefont{Alvermann,
  Edwards, and Fehske}}]{AEF07}
\bibinfo{author}{\bibfnamefont{A.}~\bibnamefont{Alvermann}},
  \bibinfo{author}{\bibfnamefont{D.~M.} \bibnamefont{Edwards}},
  \bibnamefont{and} \bibinfo{author}{\bibfnamefont{H.}~\bibnamefont{Fehske}},
  \bibinfo{journal}{Phys. Rev. Lett.} \textbf{\bibinfo{volume}{98}},
  \bibinfo{pages}{056602} (\bibinfo{year}{2007}).

\bibitem[{\citenamefont{Wellein et~al.}(2008)\citenamefont{Wellein, Fehske,
  Alvermann, and Edwards}}]{WFAE08}
\bibinfo{author}{\bibfnamefont{G.}~\bibnamefont{Wellein}},
  \bibinfo{author}{\bibfnamefont{H.}~\bibnamefont{Fehske}},
  \bibinfo{author}{\bibfnamefont{A.}~\bibnamefont{Alvermann}},
  \bibnamefont{and} \bibinfo{author}{\bibfnamefont{D.~M.}
  \bibnamefont{Edwards}}, \bibinfo{journal}{Phys. Rev. Lett.}
  \textbf{\bibinfo{volume}{101}}, \bibinfo{pages}{136402}
  (\bibinfo{year}{2008}).

\bibitem[{\citenamefont{Ejima et~al.}(2009)\citenamefont{Ejima, Hager, and
  Fehske}}]{EHF09}
\bibinfo{author}{\bibfnamefont{S.}~\bibnamefont{Ejima}},
  \bibinfo{author}{\bibfnamefont{G.}~\bibnamefont{Hager}}, \bibnamefont{and}
  \bibinfo{author}{\bibfnamefont{H.}~\bibnamefont{Fehske}},
  \bibinfo{journal}{Phys. Rev. Lett.} \textbf{\bibinfo{volume}{102}},
  \bibinfo{pages}{106404} (\bibinfo{year}{2009}).

\bibitem[{\citenamefont{Ejima and Fehske}(2009)}]{EF09b}
\bibinfo{author}{\bibfnamefont{S.}~\bibnamefont{Ejima}} \bibnamefont{and}
  \bibinfo{author}{\bibfnamefont{H.}~\bibnamefont{Fehske}},
  \bibinfo{journal}{Phys. Rev. B} \textbf{\bibinfo{volume}{80}},
  \bibinfo{pages}{155101} (\bibinfo{year}{2009}).

\bibitem[{\citenamefont{Bednorz and M\"uller}(1986)}]{BM86}
\bibinfo{author}{\bibfnamefont{I.~G.} \bibnamefont{Bednorz}} \bibnamefont{and}
  \bibinfo{author}{\bibfnamefont{K.~A.} \bibnamefont{M\"uller}},
  \bibinfo{journal}{Z. Phys. B} \textbf{\bibinfo{volume}{64}},
  \bibinfo{pages}{189} (\bibinfo{year}{1986}).

\bibitem[{\citenamefont{Dagotto}(1994)}]{Da94}
\bibinfo{author}{\bibfnamefont{E.}~\bibnamefont{Dagotto}},
  \bibinfo{journal}{Rev. Mod. Phys.} \textbf{\bibinfo{volume}{66}},
  \bibinfo{pages}{763} (\bibinfo{year}{1994}).

\bibitem[{\citenamefont{Jonker and van Santen}(1950)}]{JS50}
\bibinfo{author}{\bibfnamefont{G.~H.} \bibnamefont{Jonker}} \bibnamefont{and}
  \bibinfo{author}{\bibfnamefont{J.~H.} \bibnamefont{van Santen}},
  \bibinfo{journal}{Physica} \textbf{\bibinfo{volume}{16}},
  \bibinfo{pages}{337} (\bibinfo{year}{1950}).

\bibitem[{\citenamefont{Dagotto}(2003)}]{Da03}
\bibinfo{author}{\bibfnamefont{E.}~\bibnamefont{Dagotto}},
  \emph{\bibinfo{title}{Nanoscale Phase Separation and Colossal
  Magnetoresistance: The Physics of Manganites and Related Compounds}}
  (\bibinfo{publisher}{Springer}, \bibinfo{address}{Heidelberg},
  \bibinfo{year}{2003}).

\bibitem[{\citenamefont{Wei{\ss}e and Fehske}(2004)}]{WF04b}
\bibinfo{author}{\bibfnamefont{A.}~\bibnamefont{Wei{\ss}e}} \bibnamefont{and}
  \bibinfo{author}{\bibfnamefont{H.}~\bibnamefont{Fehske}},
  \bibinfo{journal}{New J. Phys.} \textbf{\bibinfo{volume}{6}},
  \bibinfo{pages}{158} (\bibinfo{year}{2004}).

\bibitem[{\citenamefont{Saito et~al.}(1998)\citenamefont{Saito, Dresselhaus,
  and Dresselhaus}}]{SDD98}
\bibinfo{author}{\bibfnamefont{R.}~\bibnamefont{Saito}},
  \bibinfo{author}{\bibfnamefont{G.}~\bibnamefont{Dresselhaus}},
  \bibnamefont{and} \bibinfo{author}{\bibfnamefont{M.~S.}
  \bibnamefont{Dresselhaus}}, \emph{\bibinfo{title}{Physical Properties of
  Carbon Nanotubes}} (\bibinfo{publisher}{Imperial College Press},
  \bibinfo{address}{London}, \bibinfo{year}{1998}).

\bibitem[{\citenamefont{Mousavi and Bagheri}(2012)}]{MB12}
\bibinfo{author}{\bibfnamefont{H.}~\bibnamefont{Mousavi}} \bibnamefont{and}
  \bibinfo{author}{\bibfnamefont{M.}~\bibnamefont{Bagheri}},
  \bibinfo{journal}{Physica E} \textbf{\bibinfo{volume}{44}},
  \bibinfo{pages}{1722} (\bibinfo{year}{2012}).

\bibitem[{\citenamefont{Novoselov et~al.}(2004)\citenamefont{Novoselov, Geim,
  Morozov, Jiang, Zhang, Dubonos, Grigorieva, and Firsov}}]{NGMJZDGF04}
\bibinfo{author}{\bibfnamefont{K.~S.} \bibnamefont{Novoselov}},
  \bibinfo{author}{\bibfnamefont{A.~K.} \bibnamefont{Geim}},
  \bibinfo{author}{\bibfnamefont{S.~V.} \bibnamefont{Morozov}},
  \bibinfo{author}{\bibfnamefont{D.}~\bibnamefont{Jiang}},
  \bibinfo{author}{\bibfnamefont{Y.}~\bibnamefont{Zhang}},
  \bibinfo{author}{\bibfnamefont{S.~V.} \bibnamefont{Dubonos}},
  \bibinfo{author}{\bibfnamefont{I.~V.} \bibnamefont{Grigorieva}},
  \bibnamefont{and} \bibinfo{author}{\bibfnamefont{A.~A.}
  \bibnamefont{Firsov}}, \bibinfo{journal}{Science}
  \textbf{\bibinfo{volume}{306}}, \bibinfo{pages}{666} (\bibinfo{year}{2004}).

\bibitem[{\citenamefont{Stauber and Peres}(2008)}]{SP08}
\bibinfo{author}{\bibfnamefont{T.}~\bibnamefont{Stauber}} \bibnamefont{and}
  \bibinfo{author}{\bibfnamefont{N.~M.~R.} \bibnamefont{Peres}},
  \bibinfo{journal}{J. Phys. Condens. Matter} \textbf{\bibinfo{volume}{20}},
  \bibinfo{pages}{055002} (\bibinfo{year}{2008}).

\bibitem[{\citenamefont{Clark}(1984)}]{Cl84}
\bibinfo{author}{\bibfnamefont{R.~J.~H.} \bibnamefont{Clark}}, in
  \emph{\bibinfo{booktitle}{Advances in Infrared and Raman Spectroscopy}},
  edited by \bibinfo{editor}{\bibfnamefont{R.~J.~H.} \bibnamefont{Clark}}
  \bibnamefont{and} \bibinfo{editor}{\bibfnamefont{R.~E.} \bibnamefont{Hester}}
  (\bibinfo{publisher}{Wiley Heyden}, \bibinfo{address}{New York},
  \bibinfo{year}{1984}), vol.~\bibinfo{volume}{11}, p.~\bibinfo{pages}{95}.

\bibitem[{\citenamefont{Baeriswyl and Bishop}(1987)}]{BB87}
\bibinfo{author}{\bibfnamefont{D.}~\bibnamefont{Baeriswyl}} \bibnamefont{and}
  \bibinfo{author}{\bibfnamefont{A.~R.} \bibnamefont{Bishop}},
  \bibinfo{journal}{Phys. Scr.} \textbf{\bibinfo{volume}{T19}},
  \bibinfo{pages}{239} (\bibinfo{year}{1987}).

\bibitem[{\citenamefont{White}(1992)}]{Wh92}
\bibinfo{author}{\bibfnamefont{S.~R.} \bibnamefont{White}},
  \bibinfo{journal}{Phys. Rev. Lett.} \textbf{\bibinfo{volume}{69}},
  \bibinfo{pages}{2863} (\bibinfo{year}{1992}).

\bibitem[{\citenamefont{Jeckelmann and White}(1998)}]{JW98b}
\bibinfo{author}{\bibfnamefont{E.}~\bibnamefont{Jeckelmann}} \bibnamefont{and}
  \bibinfo{author}{\bibfnamefont{S.~R.} \bibnamefont{White}},
  \bibinfo{journal}{Phys. Rev. B} \textbf{\bibinfo{volume}{57}},
  \bibinfo{pages}{6376} (\bibinfo{year}{1998}).

\bibitem[{\citenamefont{Jeckelmann}(2002)}]{Je02b}
\bibinfo{author}{\bibfnamefont{E.}~\bibnamefont{Jeckelmann}},
  \bibinfo{journal}{Phys. Rev. B} \textbf{\bibinfo{volume}{66}},
  \bibinfo{pages}{045114} (\bibinfo{year}{2002}).

\bibitem[{\citenamefont{Jeckelmann and Fehske}(2007)}]{JF07}
\bibinfo{author}{\bibfnamefont{E.}~\bibnamefont{Jeckelmann}} \bibnamefont{and}
  \bibinfo{author}{\bibfnamefont{H.}~\bibnamefont{Fehske}},
  \bibinfo{journal}{Riv. Nuovo Cimento} \textbf{\bibinfo{volume}{30}},
  \bibinfo{pages}{259} (\bibinfo{year}{2007}).

\bibitem[{\citenamefont{Becker et~al.}(2002)\citenamefont{Becker, H\"ubsch, and
  Sommer}}]{BHS02}
\bibinfo{author}{\bibfnamefont{K.~W.} \bibnamefont{Becker}},
  \bibinfo{author}{\bibfnamefont{A.}~\bibnamefont{H\"ubsch}}, \bibnamefont{and}
  \bibinfo{author}{\bibfnamefont{T.}~\bibnamefont{Sommer}},
  \bibinfo{journal}{Phys. Rev. B} \textbf{\bibinfo{volume}{66}},
  \bibinfo{pages}{235115} (\bibinfo{year}{2002});
\bibinfo{author}{\bibfnamefont{S.} \bibnamefont{Sykora}},
  \bibinfo{author}{\bibfnamefont{A.}~\bibnamefont{H\"ubsch}}, 
  \bibinfo{author}{\bibfnamefont{K.~W.} \bibnamefont{Becker}},
\bibinfo{author}{\bibfnamefont{G.}~\bibnamefont{Wellein}},\bibnamefont{and}
\bibinfo{author}{\bibfnamefont{H.}~\bibnamefont{Fehske}},
  \bibinfo{journal}{Phys. Rev. B} \textbf{\bibinfo{volume}{71}},
  \bibinfo{pages}{045112} (\bibinfo{year}{2005})

\bibitem[{\citenamefont{Sykora et~al.}(2010)\citenamefont{Sykora, Becker, and
  Fehske}}]{SBF10}
\bibinfo{author}{\bibfnamefont{S.}~\bibnamefont{Sykora}},
  \bibinfo{author}{\bibfnamefont{K.~W.} \bibnamefont{Becker}},
  \bibnamefont{and} \bibinfo{author}{\bibfnamefont{H.}~\bibnamefont{Fehske}},
  \bibinfo{journal}{Phys. Rev. B} \textbf{\bibinfo{volume}{81}},
  \bibinfo{pages}{195127} (\bibinfo{year}{2010}).

\bibitem[{\citenamefont{Bon\v{c}a et~al.}(1999)\citenamefont{Bon\v{c}a,
  Trugman, and Batisti\'{c}}}]{BTB99}
\bibinfo{author}{\bibfnamefont{J.}~\bibnamefont{Bon\v{c}a}},
  \bibinfo{author}{\bibfnamefont{S.~A.} \bibnamefont{Trugman}},
  \bibnamefont{and}
  \bibinfo{author}{\bibfnamefont{I.}~\bibnamefont{Batisti\'{c}}},
  \bibinfo{journal}{Phys. Rev. B} \textbf{\bibinfo{volume}{60}},
  \bibinfo{pages}{1633} (\bibinfo{year}{1999}).

\bibitem[{\citenamefont{Alvermann et~al.}(2010)\citenamefont{Alvermann,
  Edwards, and Fehske}}]{AEF10}
\bibinfo{author}{\bibfnamefont{A.}~\bibnamefont{Alvermann}},
  \bibinfo{author}{\bibfnamefont{D.~M.} \bibnamefont{Edwards}},
  \bibnamefont{and} \bibinfo{author}{\bibfnamefont{H.}~\bibnamefont{Fehske}},
  \bibinfo{journal}{J. Phys. Conf. Ser.} \textbf{\bibinfo{volume}{220}},
  \bibinfo{pages}{012023} (\bibinfo{year}{2010}).

\bibitem[{\citenamefont{Kane et~al.}(1989)\citenamefont{Kane, Lee, and
  Read}}]{KLR89}
\bibinfo{author}{\bibfnamefont{C.~L.} \bibnamefont{Kane}},
  \bibinfo{author}{\bibfnamefont{P.~A.} \bibnamefont{Lee}}, \bibnamefont{and}
  \bibinfo{author}{\bibfnamefont{N.}~\bibnamefont{Read}},
  \bibinfo{journal}{Phys. Rev. B} \textbf{\bibinfo{volume}{39}},
  \bibinfo{pages}{6880} (\bibinfo{year}{1989}).

\bibitem[{\citenamefont{Martinez and Horsch}(1991)}]{MH91a}
\bibinfo{author}{\bibfnamefont{G.}~\bibnamefont{Martinez}} \bibnamefont{and}
  \bibinfo{author}{\bibfnamefont{P.}~\bibnamefont{Horsch}},
  \bibinfo{journal}{Phys. Rev. B} \textbf{\bibinfo{volume}{44}},
  \bibinfo{pages}{317} (\bibinfo{year}{1991}).

\bibitem[{\citenamefont{Trugman}(1988)}]{Tr88}
\bibinfo{author}{\bibfnamefont{S.~A.} \bibnamefont{Trugman}},
  \bibinfo{journal}{Phys. Rev. B} \textbf{\bibinfo{volume}{37}},
  \bibinfo{pages}{1597} (\bibinfo{year}{1988}).

\bibitem[{\citenamefont{Emery et~al.}(1990)\citenamefont{Emery, Kivelson, and
  Lin}}]{EKL90}
\bibinfo{author}{\bibfnamefont{V.~J.} \bibnamefont{Emery}},
  \bibinfo{author}{\bibfnamefont{S.~A.} \bibnamefont{Kivelson}},
  \bibnamefont{and} \bibinfo{author}{\bibfnamefont{H.~Q.} \bibnamefont{Lin}},
  \bibinfo{journal}{Phys. Rev. Lett.} \textbf{\bibinfo{volume}{64}},
  \bibinfo{pages}{475} (\bibinfo{year}{1990}).

\bibitem[{\citenamefont{Dzierzawa}(1995)}]{Dz95}
\bibinfo{author}{\bibfnamefont{M.}~\bibnamefont{Dzierzawa}}, in
  \emph{\bibinfo{booktitle}{The Hubbard Model}}, edited by
  \bibinfo{editor}{\bibfnamefont{D.}~\bibnamefont{Baeriswyl}}
  (\bibinfo{publisher}{Plenum}, \bibinfo{address}{New York},
  \bibinfo{year}{1995}), vol. \bibinfo{volume}{343} of
  \emph{\bibinfo{series}{NATO Advanced Study Institutes, Ser. B}}.

\bibitem[{\citenamefont{Ejima et~al.}(2005)\citenamefont{Ejima, Gebhard, and
  Nishimoto}}]{EGN05}
\bibinfo{author}{\bibfnamefont{S.}~\bibnamefont{Ejima}},
  \bibinfo{author}{\bibfnamefont{F.}~\bibnamefont{Gebhard}}, \bibnamefont{and}
  \bibinfo{author}{\bibfnamefont{S.}~\bibnamefont{Nishimoto}},
  \bibinfo{journal}{Europhys. Lett.} \textbf{\bibinfo{volume}{70}},
  \bibinfo{pages}{492} (\bibinfo{year}{2005}).

\bibitem[{\citenamefont{Giamarchi}(2003)}]{Gi03}
\bibinfo{author}{\bibfnamefont{T.}~\bibnamefont{Giamarchi}},
  \emph{\bibinfo{title}{Quantum Physics in One Dimension}}
  (\bibinfo{publisher}{Clerendon Press}, \bibinfo{address}{Oxford},
  \bibinfo{year}{2003}).

\bibitem[{\citenamefont{Ogata et~al.}(1991)\citenamefont{Ogata, Luchini,
  Sorella, and Assaad}}]{OLSA91}
\bibinfo{author}{\bibfnamefont{M.}~\bibnamefont{Ogata}},
  \bibinfo{author}{\bibfnamefont{M.~U.} \bibnamefont{Luchini}},
  \bibinfo{author}{\bibfnamefont{S.}~\bibnamefont{Sorella}}, \bibnamefont{and}
  \bibinfo{author}{\bibfnamefont{F.~F.} \bibnamefont{Assaad}},
  \bibinfo{journal}{Phys. Rev. Lett.} \textbf{\bibinfo{volume}{66}},
  \bibinfo{pages}{2388} (\bibinfo{year}{1991}).

\bibitem[{\citenamefont{Moreno et~al.}(2011)\citenamefont{Moreno, Muramatsu,
  and Manmana}}]{MMM11}
\bibinfo{author}{\bibfnamefont{A.}~\bibnamefont{Moreno}},
  \bibinfo{author}{\bibfnamefont{A.}~\bibnamefont{Muramatsu}},
  \bibnamefont{and} \bibinfo{author}{\bibfnamefont{S.~R.}
  \bibnamefont{Manmana}}, \bibinfo{journal}{Phys. Rev. B}
  \textbf{\bibinfo{volume}{83}}, \bibinfo{pages}{205113}
  (\bibinfo{year}{2011}).

\bibitem[{\citenamefont{Berciu and Fehske}(2010)}]{BF10}
\bibinfo{author}{\bibfnamefont{M.}~\bibnamefont{Berciu}} \bibnamefont{and}
  \bibinfo{author}{\bibfnamefont{H.}~\bibnamefont{Fehske}},
  \bibinfo{journal}{Phys. Rev. B} \textbf{\bibinfo{volume}{82}},
  \bibinfo{pages}{085116} (\bibinfo{year}{2010}).

\bibitem[{\citenamefont{Bloch et~al.}(2008)\citenamefont{Bloch, Dalibard, and
  Zwerger}}]{BDZ08}
\bibinfo{author}{\bibfnamefont{I.}~\bibnamefont{Bloch}},
  \bibinfo{author}{\bibfnamefont{J.}~\bibnamefont{Dalibard}}, \bibnamefont{and}
  \bibinfo{author}{\bibfnamefont{W.}~\bibnamefont{Zwerger}},
  \bibinfo{journal}{Rev. Mod. Phys.} \textbf{\bibinfo{volume}{80}},
  \bibinfo{pages}{885} (\bibinfo{year}{2008}).

\bibitem[{\citenamefont{Eckardt and Lewenstein}(2010)}]{EL10}
\bibinfo{author}{\bibfnamefont{A.}~\bibnamefont{Eckardt}} \bibnamefont{and}
  \bibinfo{author}{\bibfnamefont{M.}~\bibnamefont{Lewenstein}},
  \bibinfo{journal}{Phys. Rev. A} \textbf{\bibinfo{volume}{82}},
  \bibinfo{pages}{011606} (\bibinfo{year}{2010}).

\end{thebibliography}
\bibliographystyle{apsrev}

\end{document}